\documentclass[aps,prb,preprint,a4paper,superscriptaddress,%
	notitlepage]{revtex4-1}

\usepackage{amsmath}
\usepackage{braket}
\usepackage{xspace}
\usepackage{graphicx}

\newcommand{\gsymm}{\ensuremath{\gamma^{LR}_S}\xspace}
\newcommand{\ganti}{\ensuremath{\gamma^{LR}_A}\xspace}
\newcommand{\gL}{\ensuremath{\gamma^{L}_{}}\xspace}
\newcommand{\gR}{\ensuremath{\gamma^{R}_{}}\xspace}
\newcommand{\gLexch}{\ensuremath{\gamma^L_{\mathrm{exch}}}\xspace}
\newcommand{\gLhop}{\ensuremath{\gamma^L_{\mathrm{hop}}}\xspace}

\newcommand{\gc}{\ensuremath{\gamma^{LR}_c}\xspace}
\newcommand{\gd}{\ensuremath{\gamma^{LR}_d}\xspace}

\newcommand{\bL}{\ensuremath{\mathbf{L}}\xspace}
\newcommand{\bM}{\ensuremath{\mathbf{M}}\xspace}

\begin{document}

\raggedright

\title{A universal explanation of tunneling conductance in exotic
superconductors}

\author{Jongbae Hong}
\affiliation{Center for Theoretical Physics of Complex Systems,
Institute for Basic Science, Daejeon 305-811, Korea}
\author{D.~S.~L.~Abergel$^\ast$}
\affiliation{Nordita, KTH Royal Institute of Technology and Stockholm
University, Roslagstullsbacken 23, SE-106 91 Stockholm, Sweden}
\affiliation{Center for Quantum Materials, KTH and Nordita,
Roslagstullsbacken 11, SE-106 91 Stockholm, Sweden}

\maketitle

\textbf{A longstanding mystery in understanding cuprate superconductors
is the inconsistency between the experimental data measured by scanning
tunneling spectroscopy (STS) and angle-resolved photoemission
spectroscopy (ARPES). In particular, the gap between prominent side
peaks observed in STS is much bigger than the superconducting gap
observed by ARPES measurements. Here, we reconcile the two experimental
techniques by generalising a theory which was previously applied to
zero-dimensional mesoscopic Kondo systems to strongly correlated
two-dimensional (2D) exotic superconductors.
We show that the side peaks observed in tunneling conductance
measurements in all these materials have a universal origin: They are
formed by coherence-mediated tunneling under bias and do not directly
reflect the underlying density of states (DOS) of the sample. 
We obtain theoretical predictions of the tunneling conductance and the
density of states of the sample simultaneously and show that for cuprate
and pnictide superconductors, the extracted sample DOS is consistent
with the superconducting gap measured by ARPES.}

Tunneling conductance of non-strongly-correlated materials (where the
electron--electron interactions do not have a fundamental effect on the
physics) is simply
understood because the $dI/dV$ as a function of $V$ 
(where $V$ is the voltage difference between the sample and the tip and
$I$ is the measured current) is known to correspond to the DOS of the
sample.
However, for strongly correlated materials (SCMs) this is not true
because the strong on-site interactions fundamentally change the
tunneling mechanism by which
electrons move from the current source to the sample itself.
Nevertheless, this interpretation is frequently used when discussing
SCMs, which can lead to confusing inconsistencies.
For example, a sharp peak indicating the superconducting gap of cuprate
superconductors is observed in the ARPES data near the nodal region
\cite{Lee-Nature450, Tanaka-Science314}, but in the STS data, two
prominent side peaks occur at much higher energy 
\cite{Lawler-Nature466, Pushp-Science324, Fischer-RMP79}. 
Also, STS data for strongly correlated pnictide superconductors display
two prominent side peaks which are much bigger than the energy scale of
the superconducting transition temperature \cite{Chi-PRL109}.  
Therefore, explaining the structure of tunneling conductance of SCMs is
one of the most challenging and urgent subjects in condensed matter
physics. 
This situation will remain until a comprehensive theory taking into
account the non-equilibrium nature of the STS experiment and the strong
electron--electron interactions has been developed. 

The main purpose of this study is to prove that the two prominent side
peaks in the STS data are produced by the interplay between strong
electron correlations in the sample and the non-equilibrium situation
imposed by the experimental setup.
An additional highlight of this study, just as important as the
interpretation of the side peaks, is that we can effectively find the
DOS of the specific 2D SCM under study within the fitting procedure in
the theory.
We find that the DOS predicted for cuprate and pnictide superconductors
are consistent with the data given by ARPES \cite{Tanaka-Science314,
Umezawa-PRL108}.

\begin{figure}[t]
\includegraphics[]{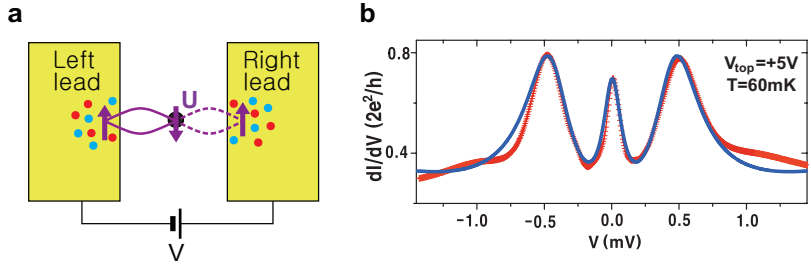}
\caption{\textbf{A schematic of the mesoscopic Kondo system and typical
experimental $dI/dV$ data.}
	\textbf{a}, An entangled state is composed of a linear combination
	of singlets (the solid and dashed magenta loops) and coherent up-
	and down-spins denoted by different colored dots. 
	\textbf{b}, Tunneling conductance for a QPC. Experimental data
	(red) reported in Ref.~\onlinecite{Sarkozy-PRB79}, theoretical results
	(blue) obtained with parameters in the first line of
	Tab.~\ref{tab:parameters}.
	\label{fig:systems}}
\end{figure}

The non-equilibrium calculation of $dI/dV$ for SCMs is achieved by
generalizing a theory for the non-equilibrium Kondo effect in
zero-dimensional mesoscopic systems \cite{Hong-JPCM23a,
Hong-JPCM23b} to the extended 2D situation. 
Some mathematical details are presented in the Methods section and in
the Supplementary Materials, but in summary, the non-equilibrium
tunneling dynamics are encoded within the non-equilibrium many-body
Green's function for the mediating site (MS, indicated by the subscript
``$d$''), $\rho_d(\omega)$, which is derived from the Liouville
operator. 
Once this is known, the tunneling conductance can be computed from the
formula in equation~\eqref{eq:dIdV}. 
The advantage of using the Liouvillian approach instead of the
Hamiltonian approach is the availability of a complete set of basis
vectors, which is not possible in the latter case. 
The difficulty with the Liouvillian approach arises in the calculation
of Liouville matrix elements for a specific model system.
This problem can be resolved by making phenomenological assumptions or
treating the unknowns as fitting parameters. 
However, for the Hamiltonian approach, one encounters a fundamental
difficulty in obtaining a complete set of basis vectors at the beginning
of the calculation and therefore progress is blocked.
The DOS in the leads are captured by the lead DOS functions $\Gamma^L$
and $\Gamma^R$ which enter via $\tilde{\Gamma}(\omega)$ and
$\rho_d(\omega)$ in equation \eqref{eq:dIdV}. Hence, these functions are
inputs to the theory, and can be chosen to represent various different
scenarios. 

\begin{figure}[t]
\includegraphics[]{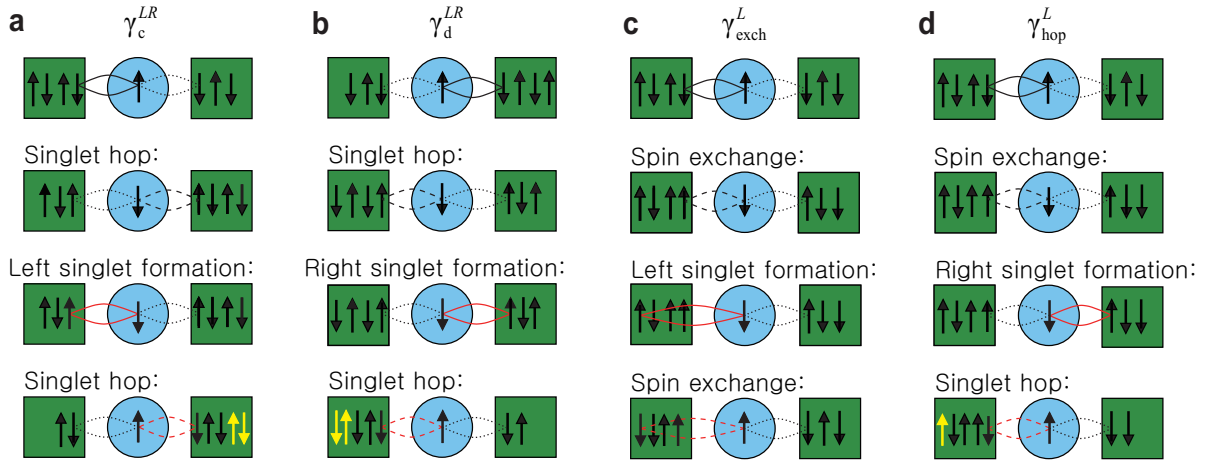}
\caption{\textbf{Coherent hybridization dynamics of $\gamma$.}
	The first row indicates the initial entangled states before
	performing hybridization dynamics (subsequent rows).  The black
	dotted line is
	added to represent an entangled state. The dashed line indicates
	the state after exchange or singlet hopping, and the red solid line
	indicates the singlet formation for next hybridization process.
	The yellow arrows indicate transferred spins.
	\textbf{a,b}, Finite-bias coherent current formation in \gc and \gd
	exclusively by singlet hopping.
	\textbf{c}, The exchange process contribution to $\gL$.
	\textbf{d}, Low-bias coherent current forming dynamics in \gL which
	is given by the combination of spin exchange and singlet hopping.
	\label{fig:dynamics}}
\end{figure}

\begin{figure}[t]
\includegraphics[]{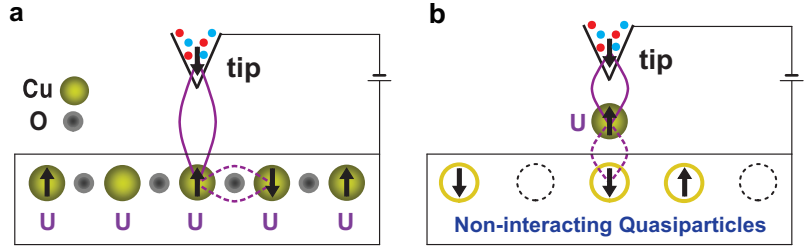}
\caption{\textbf{Schematics of STS experiments on 2D cuprate
superconductors.}
	\textbf{a}, A schematic of an STS experiment for a real cuprate
	superconductor. All Cu atoms have on-site repulsion $U$ and the MS
	is located just below the tip.
	\textbf{b} A schematic of the model from the many-body point of
	view. There is a single MS with on-site repulsion $U$, and the lower
	part is a non-interacting system of Bogoliubov quasiparticles
	excited by applied bias, whose DOS is that of a correlated
	superconductor.
	\label{fig:STSassumptions}}
\end{figure}

We now describe the coherent tunneling processes which produce the
dominant features of the experimental $dI/dV$ curves.
Our central claim is that many-body singlet states form between
electrons on the MS and in both leads (as shown by the ellipses in
figures~\ref{fig:systems}, \ref{fig:dynamics}, and
\ref{fig:STSassumptions}).
When a bias voltage is applied, the coherent singlet cotunnels
unidirectionally through the MS. 
In this context, coherence implies that the singlet may perform spin
exchange or change its partner in the leads without any energy cost. 
A mesoscopic Kondo system is sketched in figure~\ref{fig:systems}a and
consists of two leads with constant DOS and a central zero-dimensional
MS.
This system can be described by a two-reservoir Anderson
impurity model \cite{Anderson-PhysRev124}. 
To understand the coherent tunneling mechanism of the entangled state, 
we must examine the elements of the Liouville operator shown in equation
\eqref{eq:Liouville}.
These are illustrated schematically in figure~\ref{fig:dynamics} and their
full operator expressions are given in equation (S6) in the
Supplementary Information.  
To put this discussion in context, figure~\ref{fig:systems}b
shows an archetypal plot of the tunneling conductance as a function of
bias voltage $V$ for a quantum point contact (QPC) with symmetric
coupling to the leads on both sides \cite{Sarkozy-PRB79}.  Notice there
are three coherent peaks: one ``zero bias peak'', and two ``side
peaks'' which cannot be Coulomb peaks because of the small energy scale.
We shall demonstrate which of the $\gamma$ elements contribute
to the formation of these peaks.  
We first discuss the formation of the side peaks.
Figure \ref{fig:dynamics}a and its mirror image \ref{fig:dynamics}b show
coherent current formed solely from singlet hopping.
In this process, a left singlet performs a singlet hop to the right 
(second sketch) before partner exchange on the left (third sketch) 
enables a second singlet hop to the right which returns the MS to its
original state (fourth sketch). 
This coherent process allows two electrons (highlighted in yellow in the
sketch) to tunnel from the left lead to the right lead without paying
the on-site Coulomb repulsion energy cost $U$. 
Matrix elements \gsymm and \ganti in equation \eqref{eq:Liouville} are
given by symmetric and antisymmetric combinations $\gsymm=\gc+\gd$ and
$\ganti=\gc-\gd$.  
It is clear that at equilibrium (i.e., when no bias is applied across
the system), \gc and \gd will have the same
magnitude because the singlet hops will occur with equal probability in
each direction, so that $\ganti=0$ and $\gsymm \neq 0$, indicating that
$\gsymm$ contributes to the zero bias peak. 
However, out of equilibrium, either \gc or \gd will be suppressed
because it must work against the external bias in the system so that
$\gc\neq\gd$ and as a result, $\ganti\neq0$.  
This contributes an additional channel for current at finite bias, and
this manifests as the side peaks in $dI/dV$ away from equilibrium.
Therefore, it is the inclusion of left-right antisymmetric coherent
superpositions \ganti which allow us to explain the existence of the
side peaks in the tunneling conductance.
Additionally, $\gL=\gLexch + \gLhop$, where $\gLexch$
(figure~\ref{fig:dynamics}c) represents the Kondo coupling strength on the
left side, and $\gLhop$ (figure~\ref{fig:dynamics}d) describes the current
flow forming the zero bias peak, in which spin exchange is involved.
However, the zero bias peak is suppressed in the 2D SCMs because of the
interaction-induced gap at the Fermi level and so we do not focus on it
in this study.
We briefly mention that the $\gamma$ elements are not strongly
temperature dependent unless the temperature gets so high that thermal
fluctuations destroy the coherence. In fact, the main thermal effects
will be driven by the changes in the Fermi-Dirac distribution of the
leads, and changes in the sample DOS itself.

In real STS experiments for cuprate superconductors, every copper atom
has strong on-site repulsion $U$ (shown in 
figure~\ref{fig:STSassumptions}a). 
We set up our model for 2D cuprate superconductors by stating that the
tunneling current coming from the STS tip only enters through one copper
atom. Then, the tunneling mechanism depends on the strong $U$ at that
site only so that the rest of the sample can then be modeled as a system
of non-interacting Bogoliubov quasiparticles \cite{Lee-Nature450,
Yang-Nature456, Matsui-PRL90} excited by applied bias, as sketched in
figure~\ref{fig:STSassumptions}b.
Here, an ``entangled state'' comprising a linear combination of two
singlets (solid and dashed purple loops) accompanied by other coherent
spins in the tip (blue and red dots) and the sample is formed.
The Hamiltonian for the left lead is chosen to be appropriate for the
metallic tip so that $\Gamma^L$ is constant, the Hamiltonian for the
right lead is taken to be a system of Bogoliubov quasiparticles
\cite{Lee-Nature450, Yang-Nature456, Matsui-PRL90} whose
frequency-dependent DOS characterizes the superconductor. 
Therefore, the DOS of the superconductor becomes an
input to the theory [$\Gamma^R(\omega)$ in equations
\eqref{eq:Liouville} and \eqref{eq:dIdV}] and hence we can use it as a
fitting function to
distinguish between various theoretical propoals for the superconductor
DOS.
This allows us to separate the physics of the tunneling mechanism by
coherent co-tunneling of singlet states through the MS from the physics
of the superconductor in the bulk of the sample.

\begin{figure}[t]
\includegraphics[]{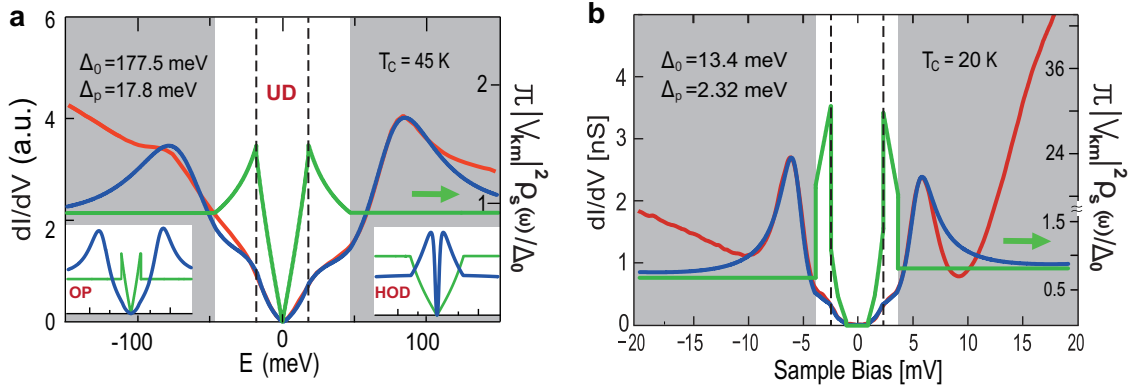}
\caption{\textbf{Comparison with experiment: Correlated superconductors.}
	In each case, the red line is experimental data, the blue line is
	the prediction of our theory, and the green line (right axis) is the
	phenomenological sample DOS used to obtain the fit.
	The vertical dashed line indicates the edge of the shoulder
	corresponding to the peak in the sample DOS. The grey shading
	denotes the region of flat sample DOS.
	\textbf{a}, STS data of cuprate superconductor
	Bi$_2$Sr$_2$CaCu$_2$O$_{8+\delta}$ given in
	Ref.~\onlinecite{Lawler-Nature466}.
	Insets: OP (left) and HOD (right) cases.
	\textbf{b},  STS data of pnictide superconductor LiFeAs given in
	Ref.~\onlinecite{Chi-PRL109}.
	\label{fig:expcomp}}
\end{figure}

Now, we apply the theory described above to correlated superconductors
to show that it gives accurate predictions for published STS data and
that the features in the sample DOS lead to the fine structure in the
low-bias region.
We require a phenomenological frequency-dependent sample DOS to use as
$\Gamma^R(\omega)$.
We find that a DOS characteristic of a $d$-wave superconducting order
parameter for the under-doped (UD) cuprate and a DOS composed of an
$s$-wave gap and a sharp DOS barrier for the pnictide give a
well-fitting tunneling conductance.
In both cases, we use flat DOS in the high frequency region to emphasize
that the side peaks are not created by features in the sample DOS.
These DOS functions are shown as the green lines (right-hand axis) in
figures~\ref{fig:expcomp}a and~\ref{fig:expcomp}b.  Choosing the
parameters given in the second and third row of
Tab.~\ref{tab:parameters} gives the fit (blue line) to the experimental
data (red line).  The quantitative agreement up to the side peaks is
almost exact, and we reproduce all the low-bias features.  The peaks of
the DOS are located at $\Delta_p=17.8\mathrm{meV}$
(figure~\ref{fig:expcomp}a) and $\Delta_p=2.3\mathrm{meV}$
(figure~\ref{fig:expcomp}b), which are consistent with the superconducting
gap reported in recent ARPES data \cite{Tanaka-Science314,Umezawa-PRL108}
and correspond to the low-energy shoulders in the $dI/dV$.
The result is natural since the Bogoliubov quasiparticles introduced in
figure~\ref{fig:STSassumptions}b describe the coherent superconducting state
\cite{Lee-Nature450}.
We also show the $dI/dV$ and the DOS for optimally doped (OP) and highly
overdoped (HOD) cases in the insets of figure~\ref{fig:expcomp}a.  It is
noteworthy that the two side peaks in the UD and OP cases are formed in
the region of flat DOS.  
This demonstrates that the side peaks are not connected to the
superconducting gap, and they must be reinterpreted as a coherent peak
of the non-equilibrium transport mechanism.  
Their sustained presence when $T>T_c$ signifies the existence
of pre-formed Cooper pairs \cite{Yang-Nature456} up to $T^\ast$ of
cuprate superconductors.  
Since a big gap observed in the anti-nodal region in the ARPES
measurements is not a coherent gap \cite{Lee-Nature450,
Tanaka-Science314}, the
anti-nodal gap is not included in this study.
A big difference occurs in the HOD case where strong
correlation no longer persists.  We obtain the HOD line shape using
vanishing $\gamma$ elements for a very weak effective Coulomb
interaction.  Note that two peaks occur within the $d$-wave region.

In conclusion, we have generalised a non-equilibrium quantum transport
theory which accurately models tunneling conductance measurements for
extended 2D strongly correlated systems. 
This theory is therefore crucial in understanding experimental data in a
wide variety of different contexts. 
A particularly interesting point is that our theoretical tunneling
conductance and sample DOS for cuprate and pnictide superconductors are
consistent with the results obtained by two different experimental
techniques.
This fact eliminates the temptation to think that the $dI/dV$ curve is
simply a reflection of the sample DOS (as it would be in the weakly
interacting case), and therefore to interpret the two side peaks as
evidence of a correlated band gap in the sample DOS. We demonstrate that
the correct interpretation is that the side peaks are evidence of the
coherent non-equilibrium tunneling mechanism.

\begin{table}[tb]
    \setlength{\tabcolsep}{5pt}
    \begin{tabular}{c|cccccccccc}
    \hline\hline
    Figure & $\gL$ & $\gR$ & $\ganti,\gsymm$ &
        $\mathrm{Re}[U^\alpha_{j^-}]$ & $\mathrm{Re}[U^L_{j^+}]$ &
        $\mathrm{Re}[U^R_{j^+}]$ & $\mathrm{Im}[U^\alpha]$ & $\Gamma^L$
		& $\Delta_0$ \\
    \hline
	\ref{fig:systems}b & 0.53 & 0.53 & 0.68 & 1.0 & 1.17  & 1.17 & 0 &
		0.8 & 0.99 \\
    \ref{fig:expcomp}a & 0.01  & 0.5 & 0.5  & 1.75  & 0.35  & 2.45 & 0.1
		& 0.24 & 118 \\
    \ref{fig:expcomp}b & 0.01  & 0.5 & 0.5  & 3.5  & 0.7 & 3.85 & 0
		& 0.24 & 13.4 
    \end{tabular}
    \vspace{0.5cm}\\
	$\beta_{11} = \beta_{15} = \beta_{55} = 0.252$,
	$\beta_{12} = \beta_{14} = \beta_{25} = \beta_{45} = 0.254$,
	$\beta_{22} = \beta_{24} = \beta_{44} = 0.258$,\\
	$\beta_{33}=1$,
	and $\beta_{13} = \beta_{23} = \beta_{43} = \beta_{53} = 0$.
    \caption{\textbf{Values of matrix elements and $\beta$.}
	The values of the matrix elements appearing in
	equation \eqref{eq:Liouville} used to create the figures. All values
	are in units of $\Delta_0$, except $\Delta_0$ ($\mathrm{meV}$).
	Note that the value in the $\Gamma^L$ column for
	figure~\ref{fig:systems}c is actually $\tilde{\Gamma}$.
    \label{tab:parameters}}
\end{table}

\section*{Methods.}
After the appropriate modelling described in the text, the STS
experiment may be described by the following Hamiltonian:
\begin{equation*}
	\mathcal{H} = \mathcal{H}^L + \mathcal{H}^{MS} + \mathcal{H}^R +
	\mathcal{H}^V
\end{equation*}
where $\mathcal{H}^L$ and $\mathcal{H}^R$ correspond to the left and right leads, which are
taken to be a metallic tip and a system of non-interacting Bogoliubov
quasiparticles (described in detail in Section III of the Supplementary
Information) excited by
the applied bias at $T=0\mathrm{K}$, respectively. The Hamiltonian for
the MS is $\mathcal{H}^{MS}$, and $\mathcal{H}^V$ describes the tunneling between the MS and
the leads. 
As described in the Supplementary Information, sections IB and IC, the
Liouville operator describing the dynamics of the MS is
\begin{equation}
    i\bL_r = \begin{pmatrix}
    0 & \gamma^{L} & -U^L_{j^-} & \gamma^{LR}_S & \gamma^{LR}_A \\
    -\gamma^{L} & 0 & -U^L_{j^+} & \gamma^{LR}_A & \gamma^{LR}_S \\
    U_{j^-}^{L*} &  U_{j^+}^{L*} & 0 &  U^{R*}_{j^+} &
U^{R*}_{j^-} \\
    -\gamma^{LR}_S & -\gamma^{LR}_A & -U_{j^+}^R  & 0 & -\gamma^{R} \\
     -\gamma^{LR}_A &  -\gamma^{LR}_S & -U_{j^-}^R & \gamma^{R} & 0
    \end{pmatrix} + i \boldsymbol{\Sigma}.
    \label{eq:Liouville}
\end{equation}
where $i\boldsymbol{\Sigma}$ denotes the self-energy matrix with elements
$i\boldsymbol{\Sigma}_{pq}(\omega) = \beta_{pq} [ \Gamma^L +
\Gamma^R(\omega)] / 2$. 
The coefficients $\beta_{pq}$ come from the process of matrix reduction
which generates equation~\eqref{eq:Liouville}. Each matrix
element has its own unique role in determining $dI/dV$. 
The middle row and column of the matrix are the incoherent couplings of
the MS to the leads which represent double occupancy at the MS via the
fluctuations of incoherent spins from the tip or the sample. 
Therefore, these elements represent the effective Coulomb interaction.
The doping effect is given by the imaginary part of these elements shown
in equation (S7). The subscript $j^{-}$ indicates
the current operator on side $\alpha$, whereas $j^+$ indicates the sum of
the left and right movements.

Within this formalism, the tunneling conductance at zero temperature can
be obtained as discussed in the Supplementary Information, sections IB
and IE:
\begin{equation}
	\frac{dI}{dV} = \left. \frac{e^2}{\hbar} \tilde{\Gamma}(\omega)
	\rho_d(\omega) \right|_{\hbar\omega=eV}
	\label{eq:dIdV}
\end{equation}
where $\tilde{\Gamma}(\omega)$ is the effective coupling of the left
lead, the MS, and the right lead and is given by
\begin{equation}
	\tilde{\Gamma}(\omega) = \frac{\Gamma^L \Gamma^R(\omega)}
	{\Gamma^L + \Gamma^R(\omega)}
	\label{eq:tildeGamma}
\end{equation}
for 2D SCMs. The quantity $\rho_d(\omega) = \frac{1}{\pi} \sum_\sigma
\mathrm{Re}[ \bM^{-1}]_{dd}$, where the matrix $\bM$ is given by the
elements $\bM_{pq} = -i\omega \delta_{p,q} + (i\bL_r)_{pq}$, is the
non-equilibrium local DOS at the MS (which is distinct from the sample
DOS) and depends on the strong interactions. It is computed from the
Liouville matrix in equation~\eqref{eq:Liouville} as described in the
Supplementary Information.


\section*{Acknowledgements.}
This work was supported by Project Code (IBS-R024-D1), the Basic Science
Research Program via the NRF Korea (2012R1A1A2005220), Nordita, ERC
project DM-321031, and partially supported by a KIAS grant funded by
MSIP. 
JH thanks Piers Coleman and Peter Fulde for helpful discussions.

\section*{Author contributions}
J.H. designed the study, performed the calculations, and wrote the
manuscript. D.S.L.A designed the study and wrote the manuscript.

\section*{Competing financial interests}
The authors declare no competing financial interests.

\end{document}